\documentstyle[aps,preprint]{revtex}
\begin{document}

\title{What can we learn from three-pion interferometry?}

\author{U. Heinz and Q.H. Zhang\cite{home}} 

\address{Institut f\"ur Theoretische Physik, Universit\"at 
Regensburg,\\
D-93040 Regensburg, Germany}

\date{\today}

\maketitle

\begin{abstract}

We address the question which additional information on the source 
shape and dynamics can be extracted from three-particle Bose-Einstein 
correlations. For chaotic sources the true three-particle correlation 
term is shown to be sensitive to the momentum dependence of the saddle 
point of the source and to its asymmetries around that point. For 
partially coherent sources the three-pion correlator allows to measure 
the degree of coherence without contamination from resonance decays. 
We derive the most general Gaussian parametrization of the two- and 
three-particle correlator for this case and discuss the space-time 
interpretation of the corresponding parameters. 

\end{abstract}


\section{Introduction}
\label{sec1}

Two-particle Bose-Einstein interferometry (also known as Hanbury 
Brown-Twiss intensity interferometry) as a method for obtaining 
information on the space-time geometry and dynamics of relativistic 
heavy ion collisions has recently received intensive theoretical and 
experimental attention. Detailed theoretical investigations (for a 
recent review see Ref.~\cite{He96}) have shown that high-quality 
two-particle correlation data can reveal not only the geometric 
extension of the particle-emitting source but also its dynamical state 
at particle freeze-out. This information is encoded in the second 
central space-time moments of the ``emission function'' $S(x,K)$, i.e.  
of the Wigner phase-space density of the source. For chaotic 
sources, certain linear combinations of these moments can be extracted 
from the two-particle correlation function $C_2(\bbox{q},\bbox{K})$ by 
fitting it to a Gaussian in the relative momentum $q$ of the pair 
\cite{BDH94,CSH95,HTWW96}. These second space-time moments give the 
size of the regions of homogeneity \cite{MS88,CSH95} which effectively 
contribute to the emission of particle pairs with a given pair 
momentum $K$; collective dynamics of the source results in a 
characteristic $K$-dependence of these homogeneity regions 
\cite{P84,WSH96,HTWW96}.  

More detailed information on the space-time structure of the source
may be hidden \cite{WH96} in possible non-Gaussian features of the 
correlation function $C_2(\bbox{q},\bbox{K})$ even if they are hard to 
extract; due to the symmetry under $q\to -q$, however, only even 
space-time moments of the source are accessible via two-particle 
correlations. In this paper we will extend previous studies of
multi-particle correlations \cite{Zajc1,Pratt3,Zhang1,Zhang3,APW93} 
and show that three-pion correlations provide in principle additional 
information on the space-time characteristics of the source which 
cannot be obtained from two-particle interferometry. We show in 
particular in Sec. \ref{sec2} that for completely chaotic sources the 
true three-pion correlations are determined by the phase of the 
two-particle exchange amplitude \cite{APW93,HV96} which drops out from 
the two-particle cross section. This phase is shown to be sensitive to 
the rate at which the saddle point $\bar x(K)$ of the source, from 
which most pairs with momentum $K$ are emitted, moves as $K$ changes, 
and to the asymmetries of the emission function around this saddle 
point via its {\it third} central space-time moments. Unfortunately, 
this phase turns out to be generically small, and its sensitivity to 
these asymmetries is very weak, making them extremely hard to measure.  

In the absence of such a non-trivial phase, three-particle correlations 
can still be used to test the chaoticity of the emitting source. To this 
end we derive the expressions for two- and three-particle correlations
for chaotic and partially coherent sources and establish their 
respective relationships. Our treatment differs from previous studies 
of multi-particle Bose-Einstein correlations in that we consistently 
express the correlation functions through the source Wigner density, 
even for partially coherent sources. This enables us to relate the 
shape of the correlators as functions of the various relative momenta 
to certain space-time features of the source. To the best of our 
knowledge the corresponding relations for partially coherent sources 
(Eqs.~(\ref{x20}) - (\ref{x22})) are new.

\section{Chaotic sources}
\label{sec2}

For a chaotic source, the two-pion correlation function $C_{2}(p_i,p_j)$ 
can be expressed as \cite{He96,APW93}
 \begin{eqnarray}
   C_{2}(p_i,p_j) &=& {P_2(p_i,p_j) \over P_1(p_i)\, P_1(p_j)} 
 \nonumber\\
   &=& 1 + {\left \vert \int d^4x \, S(x,K_{ij})\, 
                         e^{i q_{ij}\cdot x} \right \vert^2
                         \over
                         \int d^4x\, S(x,p_i) \ \int d^4y\, S(y,p_j)}
   = 1 + \frac{|\rho_{ij}|^2}{\rho_{ii} \rho_{jj}} \, .
 \label{1}
 \end{eqnarray}
Here $P_2(p_i,p_j)$ is the two-pion inclusive cross section, and 
$P_1(p_i)$ is the single-particle inclusive spectrum. $S(x,p)$ is 
the single-particle Wigner density of the source, i.e. the quantum 
mechanical analogue of its phase-space distribution.
The average and relative 4-momenta $K_{ij}=(p_i+p_j)/2$ and 
$q_{ij}=p_i-p_j$ satisfy the constraint $q_{ij}\cdot K_{ij} = 0$ 
which results from the on-shell nature of the observed momenta $p_i$. 
The two-particle exchange amplitude $\rho_{ij}$ is defined as 
\cite{APW93,CH94}
 \begin{eqnarray}
   \rho_{ij} &=& \rho(q_{ij},K_{ij}) = \sqrt{E_i\, E_j} 
      \langle \hat a^\dagger(p_i) \hat a(p_j) \rangle 
 \nonumber\\
   &=& \int  d^4x \, S(x,K_{ij}) \,
   e^{i q_{ij}\cdot x} \equiv f_{ij}\, e^{i\phi_{ij}} \, .
 \label{2}
 \end{eqnarray}
From (\ref{2}) it follows that $\rho_{ij}=\rho^*_{ji}$ and thus
$f_{ij}=f_{ji}$ and $\phi_{ij}=-\phi_{ji}$. Correspondingly, $\phi_{ii}=0$,
$\rho_{ii} = f_{ii}$, and $f_{ij}$ must be an even function of $q_{ij}$ 
while $\phi_{ij}$ is odd in $q_{ij}$.
  
The single-pion spectrum can be written as
 \begin{equation}
    P_1(p_{i}) = \int  d^4x \, S(x,p_i) = f_{ii} 
 \label{3}
 \end{equation}
while the true two-pion correlation function is defined by
 \begin{equation}
   R_2(i,j) \equiv R_2(p_i,p_j) = C_2(p_i,p_j) - 1 
   = \frac{f_{ij}^{2}}{f_{ii} \, f_{jj}} \, .
 \label{4}
 \end{equation}
Similarly, the true three-pion correlation function is given by 
\cite{APW93,W89,BBMST90,SB92,PRW92,Cramer,Eggers}
 \begin{eqnarray} 
   R_3(p_1,p_2,p_3) &=& C_3(p_1,p_2,p_3) - R_2(1,2) - R_2(2,3) - R_2(3,1) - 1
 \nonumber\\
   &=& 2 \frac{{\rm Re\, }(\rho_{12}\,\rho_{23}\,\rho_{31})}
              {f_{11}\,f_{22}\,f_{33}}
 \label{6} \\
   &=& 2 \frac{f_{12}\,f_{23}\,f_{31}}
              {f_{11}\,f_{22}\,f_{33}} \,
       \cos(\phi_{12}+\phi_{23}+\phi_{31}) \, .
 \nonumber
 \end{eqnarray}
Since the real parts $f_{ij}$ of the exchange amplitudes $\rho_{ij}$ 
can be extracted from the two-pion correlator, for chaotic sources the 
only additional information contained in the 3-pion correlation 
function resides in the phase \cite{APW93}
 \begin{equation}
 \label{6a}
   \Phi \equiv  \phi_{12}+\phi_{23}+\phi_{31} \, ;
 \end{equation}
it is a linear combination of the phases of the three exchange 
amplitudes $\rho_{12}$, $\rho_{23}$, and $\rho_{31}$ which enter
the true 3-pion correlator $R_3$. This phase is odd under interchange
of any two particles. It can be isolated by normalizing
$R_3$ with respect to the true 2-pion correlator $R_2$:
 \begin{equation}
   r_3(p_1,p_2,p_3) 
       = \frac{R_3(p_1,p_2,p_3)}{\sqrt{R_2(1,2) R_2(2,3) R_2(3,1)}}
       = 2 \cos \Phi \, .
 \label{7}
 \end{equation}

In order to understand which space-time features of the source affect 
the phase $\Phi$ (and thus the normalized true 3-pion correlation 
function $r_3$) we expand the exchange amplitude $\rho_{ij}$ for small 
values of $q_{ij} = p_i - p_j$ \cite{BDH94,CSH95}. We define the 
average of an arbitrary space-time function $f(x)$ with the source 
distribution $S(x,K_{ij})$ as
 \begin{equation}
   \langle f(x) \rangle_{ij} = \frac{\int d^4x \, f(x) \, S(x,K_{ij})}
                                    {\int d^4x \, S(x,K_{ij})} \, .
 \label{8}
 \end{equation}
This average is a function of the pair momentum $K_{ij}$. Using 
(\ref{2}) we thus get
 \begin{equation} 
   \rho_{ij} = P_1(K_{ij}) \left[
   1 + i \langle q_{ij}{\cdot}x \rangle_{ij} 
     - \frac{1}{2} \left\langle (q_{ij}{\cdot}x)^2 \right\rangle_{ij}
     - \frac{i}{6} \left\langle (q_{ij}{\cdot}x)^3 \right\rangle_{ij}
     + O\left(q_{ij}^4\right) 
                           \right] \, .
 \label{9}
 \end{equation}
Separating real and imaginary parts we find, after a little algebra,
 \begin{equation}
   f_{ij} = P_1(K_{ij}) \left[
   1 - \frac{1}{2} \left\langle (q_{ij}{\cdot}{\tilde x}_{ij})^2 
                   \right\rangle_{ij}
     + O\left(q_{ij}^4\right) 
                           \right] 
 \label{10}
 \end{equation}
and
 \begin{equation}
   \phi_{ij} = 
   q_{ij}{\cdot}\langle x \rangle_{ij} 
   - \frac{1}{6} \left\langle (q_{ij}{\cdot}{\tilde x}_{ij})^3 
                 \right\rangle_{ij}
   + O\left(q_{ij}^5\right) \, ,
 \label{11}
 \end{equation}
where
 \begin{equation}
   \tilde{x}_{ij} = x -\langle x \rangle_{ij} = x - \bar x (K_{ij})
 \label{12}
 \end{equation}
is the distance to the ``saddle point'' of the source, i.e. to the point 
of maximum emission for pions with momentum $K_{ij}$. According to 
Eqs.~(\ref{10}) and (\ref{4}), the two-pion correlator is sensitive to 
the second central (i.e. saddle-point corrected) space-time moments of
the emission function $S(x,K_{ij})$ \cite{BDH94,CSH95}, with higher 
order corrections from all even central space-time moments. The phase
$\Phi$, on the other hand, contains information on the odd space-time
moments. Expanding $S(x,K_{ij})$ around the average momentum $K$ of 
the pion triplet,
 \begin{eqnarray}
   K &=& {p_1 + p_2 + p_3 \over 3} = {K_{12} + K_{23} + K_{31} \over 3} \, ,
 \label{13}\\
   K_{ij} &=& K + {1\over 6} \left( q_{ik} + q_{jk} \right)\, , 
   \qquad i\ne j \ne k\, ,
 \label{14}
 \end{eqnarray}
and using $q_{12}+q_{23}+q_{31}=0$, we find from Eqs.~(\ref{6a}) and 
(\ref{11})
 \begin{eqnarray}
   \Phi &=& \frac{1}{2} \, q_{12}^\mu \, q_{23}^\nu 
            \left[ {\partial \langle x_\mu \rangle \over \partial K^\nu}
                 - {\partial \langle x_\nu \rangle \over \partial K^\mu}
            \right]
 \nonumber\\
        &-& {1\over 24} \left[ q_{12}^\mu q_{12}^\nu q_{23}^\lambda
                             + q_{23}^\mu q_{23}^\nu q_{12}^\lambda \right]
            \left[ {\partial^2 \langle x_\mu \rangle \over
                    \partial K^\nu \partial K^\lambda}
                 + {\partial^2 \langle x_\nu \rangle \over
                    \partial K^\lambda \partial K^\mu}
                 + {\partial^2 \langle x_\lambda \rangle \over
                    \partial K^\mu \partial K^\nu} \right]
 \label{15}\\
        &-& {1\over 2} q_{12}^\mu q_{23}^\nu (q_{12} + q_{23})^\lambda
            \, \langle \tilde x_\mu \tilde x_\nu \tilde x_\lambda \rangle
            + O(q^4) \, .
 \nonumber
 \end{eqnarray}
Here the average without subscripts
 \begin{equation}
   \langle f(x) \rangle = \frac{\int d^4x \, f(x) \, S(x,K)}
                               {\int d^4x \, S(x,K)}
 \label{16}
 \end{equation}
denotes the space-time average with the emission function evaluated at
the mean momentum $K$ of the pion triplet, and 
 \begin{equation}
   \tilde{x} = x - \langle x \rangle = x - \bar x(K)\, .
 \label{17}
 \end{equation}
Eq.~(\ref{15}) is the main new result of this Section. One easily 
checks that has it the correct symmetries under particle exchange. It 
should be noted that, due to the on-shell constraint $q_{ij}\cdot 
K_{ij}=0$, only three of the four components $q_{ij}^\mu$ are 
independent. The resulting relation 
 \begin{equation}
 \label{17a}
   (q^0)_{ij} = \bbox{q_{ij}}\cdot \bbox{\beta}_{ij}, 
   \quad \text{with}\quad  
   \bbox{\beta}_{ij} = \bbox{K}_{ij}/(K^0)_{ij}\, , 
 \end{equation}
can be used to eliminate the redundant $q$-components in 
Eq.~(\ref{15}), thereby mixing spatial and temporal components of the 
corresponding coefficients. This is a well-known problem also for the 
two-pion correlator (see, e.g., \cite{He96}) which prohibits a clean 
model-independent separation of the spatial and temporal widths of the 
source.  

Eq.~(\ref{15}) features two types of contributions to the phase 
$\Phi$: The formally leading contribution enters at second order in 
the relative momenta $q_{ij}$ and is proportional to the rate 
$\partial \bar x_\mu(K) / \partial K^\nu$ with which the saddle point 
of the emission function changes as a function of the pion momentum 
$K$. This term will in general be non-zero even for emission functions 
with a purely Gaussian $x$-dependence. It gives rise to a leading 
$q^4$-dependence of the normalized true three-particle correlator $r_3 
= 2 \cos \Phi$. At order $q^3$ the phase $\Phi$ receives additional 
contributions from the second $K$-derivatives of the saddle point as 
well as from the third central space-time moments $\langle \tilde 
x_\mu \tilde x_\nu \tilde x_\lambda \rangle$ of the source. The latter 
are the leading contributions from a possible asymmetry of the 
emission function $S(x,K)$ around its saddle point $\bar x(K)$; they 
vanish for purely Gaussian emission functions. We see that they enter 
the normalized three-particle correlator $r_3$ at order $q^5$ in a 
mixture with the $K$-dependence of the saddle point. This renders 
their isolation essentially impossible.  

In contrast to the widths of the emission function, which affect the 
two-pion correlator at {\em second} order in the relative momentum, the 
additional structural information which can (in principle) be extracted 
from the (normalized) three-pion correlator is seen to enter at 
most at {\em fourth} order in $q$. Their measurement is thus very 
sensitive to an accurate removal of all leading $q^2$-dependences by
proper normalization to the two-particle correlators. To achieve this 
looks like an extremely difficult experimental task. We are therefore
somewhat pessimistic about the short-term prospects of extracting 
additional structural information about the source from three-pion 
correlations. 

If the phase $\Phi$ and the information it contains about the source
are inaccessible, what else can three-pion correlations be used for
experimentally? The answer is that one can test the assumption that 
the source is chaotic. This has been pointed out previously in Refs.
\cite{BBMST90,PRW92} where specific simple parametrizations for the 
two- and three-particle correlators (as well as for higher order 
correlations) were assumed and the relationship between the various 
parameters was studied. We will here derive more general expressions 
which, in principle, permit such a test without making any simplifying 
assumptions about the shape of the source.

Before proceeding to the discussion of Bose-Einstein correlations 
from partially coherent sources, we would like to close this Section with 
a few short remarks on the effects from resonance decays. It is well 
known \cite{FW77,GKW79} that partial coherence in the source leads to 
incomplete correlations in the two-particle sector, in the sense that
$R_2(q,K)$ at vanishing relative momentum $q=0$ does not approach 
the ideal value $R_2(0,K)=2$ for chaotic sources. In actual experiments
there are, however, other possible reasons for apparently incomplete 
two-particle correlations. Most importantly, pions from the decay
of long-lived resonances contribute to the correlator only at very small
values of $q$ and thus (due to limited 2-track resolution) may escape
detection in the correlation signal while fully contributing to the
single-particle spectrum, thereby reducing the apparent correlation 
strength even for a completely chaotic source 
\cite{Z86,SOPW92,CLZ96,WH96a}. In a Gaussian parametrization of the 
exchange amplitude this can be implemented by writing instead of 
Eq.~(\ref{10}) for $q_{ij}\ne 0$ 
 \begin{equation}
 \label{18}
  f_{ij} = \lambda^{1/2}(K_{ij})\, P_1(K_{ij})\, 
  \exp\left[ - {1\over 2} q_{ij}^\mu q_{ij}^\nu R_{\mu\nu}(K_{ij}) 
  \right] \, ,
 \end{equation}
where, up to second order in $q$, $R^{\mu\nu}(K_{ij}) = \langle 
\tilde x^\mu_{ij} \tilde x^\nu_{ij} \rangle_{ij}$, with the source 
average on the r.h.s. being taken only over the ``core'' of pions from 
direct emission and from the decays of short-lived resonances 
\cite{He96,WH96a,C96}. The two-particle correlator then becomes 
 \begin{equation}
 \label{19}
    R_2(i,j) = \lambda(K_{ij})\, 
    {P_1^2(K_{ij}) \over P_1(p_i) \, P_1(p_j)} \, 
    \exp\left[ - q_{ij}^\mu q_{ij}^\nu R_{\mu\nu}(K_{ij}) \right] \, ,
 \end{equation}
and for vanishing relative momenta $q$ the three-particle correlation 
function assumes the value 
 \begin{equation}
 \label{20}
   C_3(p_1{=}p_2{=}p_3{=}K) = 1 + 3\, \lambda(K) + 2\, \lambda^{3/2}(K) \, .
 \end{equation}
Note, however, that the expression (\ref{7}) for the normalized true 
three-pion correlation function is not affected by resonance decay 
contributions and remains unchanged. This will no longer be true for 
partially coherent sources.  

\section{Partially coherent sources}
\label{sec3}

Expressions for the $n$-particle inclusive spectra from partially 
coherent sources have been previously derived, with differing methods, 
in Refs.~\cite{APW93,W89,BBMST90,SB92,PRW92,Cramer}. In the covariant 
current formalism of Refs.~\cite{GKW79,CH94} one decomposes the 
classical source current which creates the free pions in the final 
state into a coherent and a chaotic term:
 \begin{equation} 
   J(x) = J_{\rm coh}(x) + J_{\rm cha}(x) \, .
 \label{x5}
 \end{equation} 
Following the treatment of Ref.~\cite{CH94} this leads to the 
following definition of the single-particle Wigner density (``emission 
function'') of the source:
 \begin{eqnarray}
   S(x,K) &=& \int {d^4y\over 2 (2\pi)^3}\, e^{-iK\cdot y} 
   \langle J^*(x+{\textstyle{y\over 2}})
           J(x-{\textstyle{y\over 2}}) \rangle
 \nonumber\\
   &=& S_{\rm coh}(x,K) + S_{\rm cha}(x,K) \, ,
 \label{x6}
 \end{eqnarray} 
with
 \begin{mathletters}
 \label{x7}
 \begin{eqnarray}
 \label{x7a}
   S_{\rm coh}(x,K) &=&
   \int {d^4y\over 2 (2\pi)^3}\, e^{-iK\cdot y}\,
   J^*_{\rm coh}(x+{\textstyle{y\over 2}})
   J_{\rm coh}(x-{\textstyle{y\over 2}}) \, ,
 \\
 \label{x7b}
   S_{\rm cha}(x,K) &=& 
   \int {d^4y\over 2 (2\pi)^3}\, e^{-iK\cdot y}\,
   \langle J^*_{\rm cha}(x+{\textstyle{y\over 2}})
           J_{\rm cha}(x-{\textstyle{y\over 2}}) \rangle \, .
 \end{eqnarray}
 \end{mathletters}
The average on the r.h.s. of the definition (\ref{x7b}) for the 
chaotic part of the emission function is defined as in 
Ref.~\cite{CH94}, and we used 
 \begin{equation}
 \label{x8}
   \langle J^*_{\rm cha}(x)\, J_{\rm coh}(y) \rangle = 0 \,  .
 \end{equation}
The Wigner density of the full source is thus the sum of a coherent 
and a chaotic contribution; no mixed terms occur because the chaotic 
and coherent source currents do not interfere. This allows to carry 
over the intuitive and very successful Wigner function language for 
fully chaotic sources to the case of partially or completely coherent 
sources.  

We now write
 \begin{eqnarray}
   \rho_{ij} &=& \int d^4x\, S(x,K_{ij}) \, e^{iq_{ij}\cdot x} 
 \nonumber\\
   &=& \rho_{ij}^{\rm cha} + \rho_{ij}^{\rm coh}
   \equiv F_{ij}\, e^{i\Phi_{ij}} + f_{ij}\, e^{i\phi_{ij}} \, ,
 \label{x11}
 \end{eqnarray}
where $K_{ij} = (p_i+p_j)/2,\ q_{ij} = p_i-p_j$, and
 \begin{mathletters}
 \label{x12}
 \begin{eqnarray}
   F_{ij}\, e^{i\Phi_{ij}}
   &=& \int d^4x\, S_{\rm cha}(x,K_{ij})\, e^{iq_{ij}\cdot x} \, ,
 \label{x12b}\\
   f_{ij}\, e^{i\phi_{ij}}
   &=& \int d^4x\, S_{\rm coh}(x,K_{ij}) \, e^{iq_{ij}\cdot x} \, .
 \label{x12c}
 \end{eqnarray}
 \end{mathletters}
As shown in Ref.~\cite{APW93} this yields the two-pion correlation 
function in the form 
 \begin{equation}
   C_2(p_i,p_j) = 1 + R_2(i,j) 
                = 1+ \frac{F_{ij}^2 + 2 f_{ij} F_{ij} 
                           \cos(\Phi_{ij}-\phi_{ij})}
                          {(f_{ii}+F_{ii})(f_{jj}+F_{jj})} \, ,
 \label{x14}
 \end{equation}
while the three-particle correlation is given by
 \begin{eqnarray}
   C_3(p_1,p_2,p_3) &=& \frac{P_3(p_1,p_2,p_3)}
                             {P_1(p_1)\, P_1(p_2)\, P_1(p_3)}
   = 1 + R_2(1,2) + R_2(2,3) + R_2(3,1)
 \nonumber\\
   &+& \frac{2}{P_1(p_1)\, P_1(p_2)\, P_1(p_3)}
   \Bigl( F_{12} F_{23} F_{31} \cos(\Phi_{12}+\Phi_{23}+\Phi_{31})
 \nonumber\\
         &+& f_{12} F_{23} F_{31} \cos(\phi_{12}+\Phi_{23}+\Phi_{31})
 \nonumber\\
         &+& F_{12} f_{23} F_{31} \cos(\Phi_{12}+\phi_{23}+\Phi_{31})
 \nonumber\\
         &+& F_{12} F_{23} f_{31} \cos(\Phi_{12}+\Phi_{23}+\phi_{31})
         \Bigr) \, .
 \label{x15}
 \end{eqnarray}
Similar expressions were derived in Ref.~\cite{SB92}. The two- and 
three-particle correlations are seen to vanish for completely coherent 
sources ($F_{ij}\to 0\ \forall i,j$). In the opposite limit 
($f_{ij}\to 0\ \forall i,j$) one recovers the results from 
Sec.~\ref{sec2} for completely chaotic sources.  

The representations (\ref{x11}) and (\ref{x12}) permit us to write 
down for $F_{ij},\, f_{ij}$ and $\Phi_{ij},\, \phi_{ij}$ similar 
small-$q$ expansions as in Eqs.~(\ref{10}) and (\ref{11}); the 
corresponding averages are defined with respect to the chaotic and 
coherent parts, respectively, of the Wigner function (\ref{x6}). In 
the true two-pion correlation function $R_2(i,j)$ of Eq.~(\ref{x14}), 
the first term thus contains information on the second central 
space-time moments of $S_{\rm cha}(x,K_{ij})$ while the second term 
mixes the second moments of $S_{\rm cha}(x,K_{ij})$ and $S_{\rm 
coh}(x,K_{ij})$ in a rather nontrival way. Since the number of 
measurable parameters in $R_2(i,j)$ is the same as before, this 
implies a relative loss of information: the second space-time moments 
of $S_{\rm cha}$ and $S_{\rm coh}$ can neither be separated nor do 
they simply combine to the second central moments of the total source 
$S = S_{\rm cha} + S_{\rm coh}$.  

This complication goes hand in hand with a similar one in the 
three-pion correlator: Defining the true three-pion correlator as 
before, 
 \begin{eqnarray}
   R_3(1,2,3) &=& C_3(p_1,p_2,p_3) - 1 - R_2(1,2) - R_2(2,3) - R_2(3,1)
 \nonumber\\
   &=&\frac{2}{(f_{11}+F_{11})(f_{22}+F_{22})(f_{33}+F_{33})}
 \nonumber\\
   &&\times \Bigl(
     F_{12} F_{23} F_{31} \cos(\Phi_{12}+\Phi_{23}+\Phi_{31})
   + f_{12} F_{23} F_{31} \cos(\phi_{12}+\Phi_{23}+\Phi_{31})
 \nonumber\\
   && \ \ 
   + F_{12} f_{23} F_{31} \cos(\Phi_{12}+\phi_{23}+\Phi_{31})
   + F_{12} F_{23} f_{31} \cos(\Phi_{12}+\Phi_{23}+\phi_{31})
   \Bigr) ,
 \label{x16}
 \end{eqnarray}
one sees that, in contrast to Eq.~(\ref{7}) for chaotic sources, the 
phase factors can no longer be isolated by normalizing $R_3$ with a 
proper combination of two-particle correlators $R_2$. This means that,
in a samll-$q$ expansion, $R_3(1,2,3)$ contains leading terms of 
second order in $q$ which are independent of those occurring in the 
two-particle correlator. On the one hand, those terms supplement the 
incomplete information from $R_2$ on the second space-time moments of 
the source; on the other hand, they render the measurements of source 
asymmetries impossible.  

The full reconstruction of all the (in principle) measurable information 
obviously requires a measurement of $R_2(i,j)$ and $R_3(1,2,3)$ as a 
function of all nine components of $\bbox{p}_1,\bbox{p}_2,\bbox{p}_3$ . 
In view of the technical complexity (both experimental and 
theoretical) of such a program this is not likely to happen soon. It 
must, however, be mentioned that simple one- or two-parameter Gaussian 
parametrizations as suggested in Refs.~\cite{BBMST90,PRW92,HV96} are 
not sufficient for this purpose because they very strongly prejudice 
the form of the source.  

To pursue this last point a little further, let us define the 
(momentum-dependent) chaotic fraction of the single particle spectrum 
 \begin{equation}
   \epsilon(p_i)=\frac{F_{ii}}{f_{ii}+F_{ii}}
   = \frac{\int d^4x\, S_{\rm cha}(x,p_i)}{\int d^4x\, S(x,p_i)}
 \label{x17}
 \end{equation} 
The coherent fraction is accordingly $f_{ii}/(f_{ii}+F_{ii})=1-
\epsilon(p_i)$. For vanishing relative momentum 
$q_{ij}=0 \ (i,j=1,2,3)$, we then have 
 \begin{eqnarray}
   R_2(p,p) &=& \epsilon(p)\bigl(2-\epsilon(p)\bigr)\, ,
 \nonumber\\
   R_3(p,p,p) &=& 2\, \epsilon^2(p) \, \bigl(3-2\epsilon(p)\bigr) \,.
 \label{x18}
 \end{eqnarray}
For completely chaotic sources, $\epsilon(p) = 1$, we recover the results 
of Sec.~\ref{sec2}. For partially coherent sources, the normalized 
three pion correlator $r_3$ at vanishing $q$ is given by 
 \begin{equation}
   r_3(p,p,p) = \frac{R_3(p,p,p)}{\bigl(R_2(p,p)\bigr)^{3/2}}
              = 2 \sqrt{\epsilon(p)} \frac{(3-2\epsilon(p))}
                                          {(2-\epsilon(p))^{3/2}}
 \label{x19}
 \end{equation}
which, in general, deviates from the chaotic limit $r_3(p,p,p)=2$.

It would thus seem to be a simple matter to check the limits of $R_2$ 
and $R_3$ for vanishing relative momenta and construct the ratio 
(\ref{x19}) in order to see whether or not the source contains a 
coherent component. In practice, however, the $q=0$ limit can not be 
measured directly, but requires an extrapolation of data at finite $q$ 
to zero relative momenta. It is well known that such an extroplation 
can be very sensitive to the assumed functional behavior of the 
correlator at small $q$. As we will now show our results provide a 
basis for a reasonable parametrization of $R_2$ and $R_3$ for small 
$q$.  

To this end we start from Eqs.~(\ref{x14}) and (\ref{x16}) together 
with the small $q$ expansions (\ref{10}), (\ref{11}). Noting that 
$R_2$ must vanish for $q\to \infty$, a parametrization which 
is correct up to second order in $q$ is given by 
 \begin{eqnarray}
   R_2(i,j) &\approx& \epsilon^2(K_{ij})\, 
      \exp\left[ - q_{ij}^\mu q_{ij}^\nu R_{\mu\nu}(K_{ij}) \right]
\nonumber\\
   &+& 2 \epsilon(K_{ij})(1-\epsilon(K_{ij}))
      \exp\left[ -{\textstyle{1\over 2}} q_{ij}^{\mu} q_{ij}^{\nu}
      \left(R_{\mu\nu}(K_{ij}) + r_{\mu\nu}(K_{ij})\right) \right]
      \cos\left(q_{ij}{\cdot}s(K_{ij})\right)\, .
 \label{x20}
 \end{eqnarray}
It follows from Eqs.~(\ref{10}), (\ref{11}) that here
 \begin{mathletters}
 \label{x21}
 \begin{eqnarray}
   R^{\mu\nu}(K_{ij}) 
   &=& \langle \tilde{x}_{ij}^\mu \tilde{x}_{ij}^\nu \rangle_{ij}^{\rm cha}
   \, ,
 \label{x21a}\\
   r^{\mu\nu}(K_{ij})
   &=& \langle \tilde{x}_{ij}^\mu \tilde{x}_{ij}^\nu \rangle_{ij}^{\rm coh}
   \, ,
 \label{x21b}\\
   s^{\mu}(K_{ij})
   &=& \langle x^\mu \rangle^{\rm cha}_{ij}
     - \langle x^\mu \rangle^{\rm coh}_{ij} 
   \, .
 \label{x21c}
 \end{eqnarray}
 \end{mathletters}
Eq.~(\ref{x20}) neglects an additional factor $P^2(K_{ij})/P(p_i) 
P(p_j)$ which is unity for exponentical single particle spectra 
\cite{CSH95}. Eq.~(\ref{x20}) differs from the parametrization 
suggested in Ref.~\cite{PRW92} by the factor $\cos\left(q_{ij}\cdot 
s(K)\right)\, \exp\left[ - {1\over 2} q_{ij}^{\mu} q_{ij}^\nu 
r_{\mu\nu}(K_{ij}) \right]$; the parametrization of Ref.~\cite{PRW92} 
is thus not general enough. (It essentially assumes that the coherent 
part of the source is pointlike (in space {\em and} time!) and 
localized at the saddle point of the chaotic part of the source.) Note 
that from Eq.~(\ref{x20}) one must still eliminate the redundant 
$q$-component via the on-shell constraint (\ref{17a}).  

The three-pion correlator can similarly parametrized as 
 \begin{eqnarray}
   R_3(p_1,p_2,p_3) &=& 2 \epsilon^2(K)
   \exp \left[ - \left( q_{12}^\mu q_{12}^\nu + q_{23}^\mu q_{23}^\nu
      + {\textstyle{1\over 2}} (q_{12}^\mu q_{23}^\nu + q_{12}^\mu q_{23}^\nu)
                 \right) R_{\mu\nu}(K) \right]
 \nonumber\\
   &\times& \Bigl[\ \ \epsilon(K)
 \nonumber\\
   && \ + (1-\epsilon(K))\, \cos(q_{12}{\cdot}s(K))\,
          \exp\left( {\textstyle{1\over 2}} q_{12}^\mu q_{12}^\nu
                     (R_{\mu\nu}(K)-r_{\mu\nu}(K))\right)
 \nonumber\\
   && \ + (1-\epsilon(K))\, \cos(q_{23}{\cdot}s(K))\,
          \exp\left( {\textstyle{1\over 2}} q_{23}^\mu q_{23}^\nu
                     (R_{\mu\nu}(K)-r_{\mu\nu}(K))\right)
 \nonumber\\
   && \ + (1-\epsilon(K))\, \cos((q_{12}+q_{31}){\cdot}s(K))
 \nonumber\\
   &&\qquad \times \exp\left( {\textstyle{1\over 2}}(q_{12}+q_{23})^\mu 
            (q_{12}+q_{23})^\nu (R_{\mu\nu}(K)-r_{\mu\nu}(K))
                       \right) \Bigr] \, .
 \label{x22}
 \end{eqnarray}
This again generalizes the parametrizations given in 
Refs.~\cite{BBMST90,PRW92}; according to Eqs.~(\ref{10}), (\ref{11}),
it is correct up to the second order in $q$ if one approximates  
$P_1^2(K_{ij}/P_1(p_i) /P_1(p_j) \approx 1$ as well as 
$\epsilon(K_{ij}) \approx \epsilon(K)$. The parametrizations of 
Ref.~\cite{BBMST90,PRW92} are recovered in the limit of a pointlike 
coherent source, $r_{\mu\nu}(K)=0$, and assuming $s(K)=0$. (The first 
of these two assumptions is explicity stated in Ref\cite{BBMST90}.) 
One can easily convince oneself that at $q_{12}=0$, for example, the 
term $\cos(q_{23}{\cdot}s(K))\, \exp\left[\frac{1}{2}q_{23}^\mu 
q_{23}^\nu \left( R_{\mu\nu}(K)-r_{\mu\nu}(K) \right) \right]$ enters 
$R_3(q_{23})$ with a different weight than $R_2(q_{23})$. Thus $R_{3}$ 
provides additional information which allows to separate 
$R_{\mu\nu}(K)$ from $r_{\mu\nu}(K)$ and thereby the widths of the 
chaotic and coherent parts of the source.  

In practice, one must also take into account resonance decays. Since 
it follows from the discussion at the end of in Sec.~\ref{sec2} that 
the longlived resonances do not affect the intercept (\ref{x19}) of 
the normalized true three-pion correlator, and it was shown in 
Refs.~\cite{He96,WH96a,C96} that expression (\ref{x21a}) remains 
essentially valid if the chaotic part of the emission function is 
restricted to the ``core'' of direct pions and short-lived resonance 
decays, we expect Eqs.~(\ref{x20}) - (\ref{x22}) to be practically 
useful even when resonance decays are included.

\section{Conclusions}
\label{sec4}

We have studied the question to what extent three-pion Bose-Einstein 
correlations can provide independent information about the space-time 
structure of the emitting source which cannot be extracted from 
two-pion correlations. For chaotic sources we found that the 
three-pion correlator depends on the phase of the two-particle 
exchange amplitude which drops out from the two-particle cross 
section. This phase can be isolated by proper normalization of the 
true three-pion correlator with respect to the two-pion correlator.
It was shown to be sensitive to the momentum dependence of
the point of highest emissivity in the source and to the asymmetries of 
the emission function around that point. However, this sensitivity is 
weak (it enters only at 4th order in the relative momenta $q_{ij}$),
and the corresponding source properties are hard to measure.

We then proceeded to study sources which are not completely chaotic 
but contain a coherent component. We showed that in this case the 
emission function can be written as a sum of two Wigner densities
describing the chaotic and coherent components, respectively, and 
expressed the two- and three-pion correlation functions via these 
chaotic and coherent Wigner densities. We showed that a comparison of 
two- and three-pion correlators allows for a determination of the 
degree of coherence in the source, without contaminations from 
resonance decays. To this end one must study the respective 
correlation functions at vanishing relative momenta of all particles. 
To facilitate the extraction of this limit from experimental data we 
derived in Eqs.~(\ref{x20}) and (\ref{x22}) the most general 
parametrizations for the two-and three-pion correlation functions at 
small relative momenta. These new parametrizations are based on our 
expressions of the correlation functions in terms of the Wigner 
density of the source; they are exact up to second order in the 
relative momenta, i.e. for emission functions $S(x,K)$ with a Gaussian 
$x$-dependence. After eliminating the redundant $q$-components, they 
are seen to depend on 16 parameters which are all functions of the 
average momentum $\bbox{K}$ of the pion pair resp. triplet. To 
determine all these parameter functions, a complete study of the 
two-and three-particle spectra as functions of all 6 + 9 = 15 momentum 
components is necessary. (The 16th parameter, $\epsilon(K)$, describes 
the degree of coherence and enters the normalization of the 
correlation functions at vanishing relative momenta.) This is 
certainly not an easy task, and it might be worthwhile to study 
whether, for certain simple but not too unrealistic models for the 
emission function, it is not possible to obtain simpler 
parametrizations (for example by exploiting certain symmetries of the 
source).  

Our results show that in the case of partially coherent sources the 
three-pion correlator contains independent information on the second 
space-time moments of the source which cannot be extracted from the 
two-pion correlator. This information is needed to separate the 
space-time characteristics (lengths of homogeneity or effective 
widths) of the chaotic and coherent parts of the emission function. To 
extract it in practice will not be easy, but the theoretical framework 
by which this should be done has been presented here.

\acknowledgements

The authors would like to express their gratitude to T. Cs\"org\H o, 
H. Heiselberg, A.P. Vischer, and U.A. Wiedemann for stimulating 
remarks. This work arose from discussions at the Workshop on Particle 
Interferometry in High Energy Heavy Ion Reactions (HBT96) at the ECT* in 
Trento, Sept. 16 - 27, 1996. We would like to thank the ECT* for their 
hospitality and for providing such a fruitful and stimulating 
atmosphere. Q.H.Z.  gratefully acknowledges support by the Alexander 
von Humboldt Foundation through a Research Fellowship. The work of 
U.H. was supported in part by BMBF, DFG, and GSI.

\end{document}